\documentstyle[epsfig,12pt]{article} 
\begin{document} 
\begin{center} 
{\Large {\bf Spin correlations  
and consequences of quantum--mechanical coherence}} 
\end{center} 
\vskip 0.15in 
\begin{center} 
{\rm 
R.~Lednick\'{y}}$^{a,b}$, V.L.~Lyuboshitz$^{a}$ \\ 
\end{center}
\vskip 0.15in
\begin{center}
 $^a$
{\it Joint Institute for Nuclear Research, 
Dubna, Moscow Region 141980, Russia}\\
 $^b$
{\it Institute of Physics ASCR, 
Na Slovance 2, 18221 Prague 8, Czech Republic} \\ 
\vskip 0.05in 
e-mail: lednicky@fzu.cz, lyubosh@sunhe.jinr.ru\\
\end{center}
 
\begin{abstract} 
The difference in the properties of the 
spin correlation tensor for
factorizable and nonfactorizable two-particle
states is analyzed. 
The inequalities for linear combinations of the components of this 
tensor are obtained for the case of 
incoherent mixtures of factorizable two-particle
spin states.
They include the well known Bell inequalities and can be violated for 
coherent superpositions of two--particle spin states.
The possibility to  verify
the consequences of the quantum-mechanical coherence is discussed
using the angular correlations in the asymmetric (parity violating)
decays of the pairs of spin-1/2 particles (muons,
top-quarks or $\Lambda$-hyperons),
the coherence arising either from the production dynamics
or due to the effect of quantum statistics.
\end{abstract} 

\section{Introduction } 
It is well known that correlations in the detection
of nonfactorizable two-particle states represent a manifestation of the
quantum-mechanical effect first considered by Einstein,
Podolsky and Rosen \cite{l1}. The essence of this effect is
as follows. If the two-particle state is not factorizable,
the character  of the measurements performed on the first
particle determines the readout of the detector 
analyzing the state of the second particle, even if
the detectors were situated at a large distance.
To demonstrate this effect, consider
a nonfactorizable two-particle state as a coherent superposition
of pairs of one-particle states:
\begin{equation}
\label{eq1}
\mid \Phi \rangle^{(1,2)} =\sum_i \sum_k c_{ik} \mid i
\rangle^{(1)}\mid k \rangle^{(2)},
\end{equation}
where $c_{ik}$  are complex numbers normalized to unity,
$
\sum_i\sum_k\mid c_{ik}\mid^2 =1.
$
In this case, the amplitude to observe 
the two-particle  state (\ref{eq1}) by two
one-particle detectors selecting the states $\mid L
\rangle^{(1)}$ and $\mid M \rangle^{(2)}$ results from
the interference of pairs of one-particle states:
$
 A_{LM} = \sum_i\sum_k c_{ik} \langle L
\mid i \rangle^{(1)} \langle M \mid k \rangle^{(2)}.
$
Clearly, the
selection of different states $\mid L\rangle^{(1)}$ and
$\mid M \rangle^{(1)}$ for the first particle then leads to
the different states of the second particle:
\begin{equation}
\label{eq4}
 \mid \Psi \rangle_L^{(2)} = \sum_i \sum_k
c_{ik} \langle L \mid i \rangle \mid k \rangle^{(2)},\qquad
\mid \Psi \rangle_M^{(2)} = \sum_i \sum_k c_{ik} \langle M
\mid i \rangle \mid k \rangle^{(2)}.
\end{equation}

Let us note that the states $\mid\Psi \rangle_L^{(2)}$  and
$\mid\Psi \rangle_M^{(2)}$
can be the eigenfunctions of noncommuting operators. As a result, in
the presence of the correlations, the one-particle state is not 
a pure one -
it should be described by the density matrix and not by the wave
function. 

The fact that the state of one of two particles can be managed
without a direct force action on it, Einstein
considered as a paradox pointing to the
incompleteness of the quantum-mechanical description \cite{l1}.
However, it has become clear that here we deal with the
correlation effect connected with coherent properties of
quantum-mechanical superpositions. The properties of
$K^0\overline K^0$-pairs provide an impressive example:
the detection of one of two neutral kaons through its decay
or its interaction determines the internal state of the
second kaon [2-5]. 

The polarization correlations, discussed in present paper,
belong to the same group of phenomena 
[6-8].
It should be
emphasized that namely in these cases the so-called Bell
inequalities are violated. These inequalities were derived
at the probability level without taking into account the
coherent properties of the quantum-mechanical superpositions \cite{l10,l12}.
We prove here a class of the inequalities, including those of Bell, 
based on
the assumption of the factorizability of the two-particle density matrix, 
i.e. on its reduction to a sum of the direct products of one--particle density
matrices with the nonnegative coefficients.
Clearly, such a form of the density matrix corresponds to a classical
probabilistic description and cannot account for the 
coherent quantum--mechanical effects.
The violation of the Bell--type inequalities thus clearly 
manifests the coherent
nature of quantum mechanics.
We show here that the correlations in the parity-violating 
decays of pairs of 
spin-1/2 particles 
can serve as a sensitive and relatively simple test of this coherence,
extending the tests already done with optical photons and secondary
scatterings of low energy protons (see, e.g., a review \cite{l12}).

\section{Two-particle density matrix and spin correlations}

For two spin-1/2 particles, the normalized spin density matrix, 
with the sum of the diagonal elements ( "trace")
$
{\rm tr}_{(1,2)}\hat{\rho}^{(1,2)} = 1
$,
has the following general structure (see, e.g., \cite{l7})
\begin{equation}
\label{eq6}
\hat{\rho}^{(1,2)} =\frac{1}{4} [\hat{I}^{(1)} \otimes
\hat{I}^{(2)} + (\hat{\mbox{\boldmath $\sigma$}}^{(1)} {\bf P}_1)
\otimes \hat{I}^{(2)} + \hat{I}^{(1)} \otimes
(\hat{\mbox{\boldmath $\sigma$}}^{(2)} {\bf P}_2) + \sum_{i=1}^3
\sum_{k=1}^3 T_{ik}\hat{\sigma}_i^{(1)} \otimes
\hat{\sigma}_k^{(2)}].
\end{equation}
Here $ \hat{I}$ is the two-row unit matrix, 
$\hat{\mbox{\boldmath $\sigma$}}$ is the Pauli vector operator,
$ T_{ik}= \langle
\hat{\sigma}_i^{(1)} \otimes \hat{\sigma}_k^{(2)}\rangle $
are the components of the correlation tensor. 
The corresponding one-particle density matrices contain
the polarization vectors 
${\bf P}_l= \langle \hat{\mbox{\boldmath $\sigma$}}^{(l)}\rangle$ only:
$
\hat{\rho}^{(l)} = \frac{1}{2} (\hat{I} +
\hat{\mbox{\boldmath $\sigma$}} \vec{P}_l),~l=1,2.
$
In the absence of correlations the factorization takes place:
\begin{equation}
\label{eq8}
T_{ik} = P_{1i}P_{2k}, \qquad \hat{\rho}^{(1,2)} =
\hat{\rho}^{(1)} \otimes \hat{\rho}^{(2)}.
\end{equation}

Let two analyzers select the states of the first and the second
particle with the polarization vectors $\mbox{\boldmath $\zeta$}^{(1)}$
and $\mbox{\boldmath $\zeta$}^{(2)}$.  Then the detection probability $W$
depends linearly on the polarization parameters of the
two-particle system as well as on the final polarization
parameters fixed by detectors, and it can be obtained by
the substitution of the matrices $\hat{\sigma}_i^{(1)}$
and $\hat{\sigma}_k^{(2)}$ in the expression (6) with the corresponding
 projections $\zeta_i^{(1)}$ and $\zeta_k^{(2)}$.
 As a result \cite{l7}
\begin{equation}
\label{eq9}
W  \sim 1 + {\bf P}_1 \mbox{\boldmath $\zeta$}^{(1)} +
{\bf P}_2 \mbox{\boldmath $\zeta$}^{(2)} + \sum_{i=1}^3 \sum_{k=1}^3
T_{ik} \zeta_i^{(1)} \zeta_k^{(2)} .
\end{equation}
Let only the polarization vector $\mbox{\boldmath $\zeta$}^{(1)}$   of
the first particle is measured. Then, due to the
correlations, the spin state of the second particle
is described by the
normalized density matrix
\begin{equation}
\label{eq10}
\hat{\tilde{\rho}}^{(2)} = \frac{1}{2} ( 1 +
\mbox{\boldmath $\zeta$}^{(1)} {\bf P}_1 )^{-1} [ ( 1 +
\mbox{\boldmath $\zeta$}^{(1)} {\bf P}_1) \hat{I} + 
\hat{\mbox{\boldmath $\sigma$}}
{\bf P}_2 + \sum_{i=1}^3 \sum_{k=1}^3 T_{ik} \zeta_i^{(1)}
\hat{\sigma}_k].
\end{equation}
In this case the polarization vector of the second
particle has the components
\begin{equation}
\label{eq11}
\tilde{\zeta}_k^{(2)}=( P_{2k} + \sum_{i=1}^3 T_{ik}
\zeta_i^{(1)}) /  ( 1 + \mbox{\boldmath $\zeta$}^{(1)} {\bf P}_1 ).
\end{equation}
In the case of independent particles, when the factorization takes
place (see Eq. (\ref{eq8}), 
the detection of the spin state of the first particle
does not influence the polarization of the second particle:
$\tilde{\mbox{\boldmath $\zeta$}}^{(2)} = {\bf P}_2 $.

The situation is of interest when both 
one-particle states are unpolarized and
the polarization
vectors ${\bf P}_1$ and $ {\bf P}_2$ vanish.
Then, in accordance with Eq. (\ref{eq11}), the spin effects are
completely determined by the correlation tensor $T_{ik}$:
$
\tilde{\zeta}_k^{(2)} = \sum_{i=1}^3 T_{ik} \zeta _i^{(1)}.
$
If the one-particle states are unpolarized and the spin correlations
are absent, then $\tilde{\mbox{\boldmath $\zeta$}}^{(2)} = 0 $  for
any selection of the vector $\mbox{\boldmath $\zeta$}^{(1)}$.

It may be useful to calculate the polarization
vectors ${\bf P}_l=\langle\Psi_{SM}|\hat{\mbox{\boldmath $\sigma$}}^{(l)}| 
\Psi_{SM}\rangle$ and the correlation tensor
$T_{ik}=\langle\Psi_{SM}|\hat{\sigma}^{(1)}_i\otimes
\hat{\sigma}^{(2)}_k
|\Psi_{SM}\rangle$
in the pure singlet ($S=0$) and triplet ($S=1$) states $|\Psi_{SM}\rangle$.

The singlet state of
two spin-1/2 particles is a typical example of a nonfactorizable
two-particle state. It is described by the spin wave function 
\begin{equation} 
\label{eq15}
\mid \Psi \rangle_{00} =
\frac{1}{\sqrt{2}} ( \mid + 1/2 \rangle_z^{(1)} \mid -1/2
\rangle_z^{(2)} - \mid -1/2 \rangle_z^{(1)} \mid +1/2
\rangle_z^{(2)}),
\end{equation}
corresponding to rigidly correlated particle spins with the
spin projections opposite for any choice of
the quantization axis $z$; at the same time, 
the particle polarizations are equal to zero: 
\begin{equation}
\label{eq23}
{\bf P}_1 = {\bf P}_2 =0, \qquad   T_{ik} = - \delta_{ik}.
\end{equation}

Consider further the triplet states 
which are polarized and aligned along the spin axis unit vector
${\bf e}$, so that the triplet density matrix is diagonal
in the representation of the spin projections onto the axis ${\bf e}$:
$\rho_{mm'}=w_m\delta_{mm'}$. 
The states with the spin
projections $M_{\bf e}= +1, -1, 0$
can be respectively written as
\begin{equation}
\label{eq25}
\begin{array}{c}
\mid \Psi \rangle_{1+1} = \mid +1/2 \rangle_{{\bf e}}^{(1)}
 \mid +1/2 \rangle_{{\bf e}}^{(2)}, \qquad \mid \Psi \rangle_{1-1} =
\mid -1/2 \rangle_{{\bf e}}^{(1)} \mid -1/2 \rangle_{{\bf e}}^{(2)} ,
\\ \\
\mid \Psi \rangle_{10} = \frac{1}{\sqrt{2}} ( \mid +1/2
\rangle_{{\bf e}}^{(1)} \mid -1/2 \rangle_{{\bf e}}^{(2)} + \mid -1/2
\rangle_{{\bf e}}^{(1)} \mid +1/2 \rangle_{{\bf e}}^{(2)}).
\end{array}
\end{equation}
Normalizing the corresponding occupancies to unity:
$ w_+  +  w_- +  w_0   = 1$,
one has:
\begin{equation}
\label{eq29}
{\bf P}_1 = {\bf P}_2 = ( w_+ - w_- ){\bf e},~~
T_{ik} = (1 - 3w_0)e_i e_k + w_0 \delta_{ik} .
\end{equation}
In case of the unpolarized triplet
$ w_+ = w_- = w_0  = 1/3$, so that 
$
{\bf P}_1 = {\bf P}_2 = 0,~T_{ik} = \delta_{ik}/3 .
$
Note that for a general triplet density matrix $\rho_{mm'}$:
\begin{equation}
\begin{array}{c}
\label{eq29a}
P_1 = \sqrt{2}\Re (\rho_{+0}+\rho_{-0}),~
P_2 = -\sqrt{2}\Im (\rho_{+0}-\rho_{-0}),~
P_3 = \rho_{++}-\rho_{--},
\\ \\
T_{11} = \rho_{00}+2\Re\rho_{+-},~
T_{22} = \rho_{00}-2\Re\rho_{+-},~
T_{33} = \rho_{++}+\rho_{--}-\rho_{00}=1-2\rho_{00},~
\\ \\
T_{12} = T_{21}=-2\Im\rho_{+-},~
T_{13} = T_{31}=\sqrt{2}\Re(\rho_{+0}-\rho_{-0}),~
T_{23} = T_{32}=-\sqrt{2}\Im(\rho_{+0}+\rho_{-0}).
\end{array}
\end{equation}
One may see that
the diagonal components of the triplet correlation tensor
always satisfy the equalities ($w_0\equiv \rho_{00}$):
\begin{equation}
\label{eq29b}
T_{11}+T_{22} = 2w_0,~~T_{33}=1-2w_0,~~{\rm tr}T=1.
\end{equation}

\section{Analyzers of the spin polarization}
As for the polarization analyzers, one can use the secondary
scattering events and exploit the fact that
the scattering of a particle with spin 1/2 on a
spinless or unpolarized target selects the states polarized
parallel to the normal of the scattering plane.
The final polarization vectors in Eq. (\ref{eq9})
are then the analyzing powers:
$
\mbox{\boldmath $\zeta$}^{(l)} = \alpha_l({\bf p}_l,\theta_l){\bf n}^{(l)},
~ l=1,2.
$
Here ${\bf p}_l$ are the three-momenta of the produced particles,
$\theta_l$ are the scattering angles,
${\bf n}^{(l)}$ are the unit vectors parallel to
the scattering plane normals and $\alpha_l$
are the left-right azimuthal asymmetry
parameters vanishing at zero scattering angle.
According to the Wolfenstein theorem \cite{l13},
the analyzing power coincides with the
polarization vector arising as a result of the elastic scattering of
the unpolarized particle on the same target.  
Inserting the analyzing powers $\mbox{\boldmath $\zeta$}^{(l)}$ in
Eq. (\ref{eq9}), one obtains the probability of the simultaneous
detection of two particles, produced in the same collision
and subsequently scattered on an unpolarized target,
describing the correlation of the scattering planes \cite{l7}.

It is interesting to note that if two unpolarized particles
are produced and only
one of them is scattered on an
unpolarized target, then the spin correlation results in the polarization
of the other (unscattered) particle created together with the
scattered one in the same collision event (see Eq. (\ref{eq11})):
$
\tilde{\zeta}_k^{(2)}= \alpha_1({\bf p}_1,\theta_1) \sum_{i=1}^3
T_{ik}n_i^{(1)}.
$
This phenomenon makes it possible, in principle, to prepare particle beams
with regulated spin polarization without any direct action on the
particles in the polarized beam.

Compared to the secondary scattering, there is 
often a more easy way to analyze the
spin states of produced spin--1/2 particles
using their asymmetric (parity violating) decays
(see, e.g., \cite{l15,l16}).
The parity violation is characterized by the asymmetry parameter
$\alpha$ measuring the strength of the correlation between the 
particle polarization ${\bf P}$ and the decay analyzer
unit vector ${\bf n}$.
 
For example, $\alpha =0.642$ for the decay
$\Lambda \rightarrow p \pi^- $ with the decay analyzer chosen
parallel to the proton momentum in the $\Lambda$-rest frame.
Other examples are the muon decay:
$\mu^- \rightarrow e^-\bar{\nu}_e\nu_{\mu}$
with the asymmetry parameter $\alpha_e=-1/3$,
and the top--quark decay:
$t \rightarrow b l^+ \nu_l~(b \bar{d} u,~b \bar{s} c)$
with the asymmetry parameters \cite{l16a}
$\alpha_b\doteq -0.4$, $\alpha_{\nu_l (u,c)}\doteq -0.33$ and
$\alpha_{l^+ (\bar{d},\bar{s})}=1$.
Due to the CP invariance, the asymmetry parameters for the decays
of corresponding antiparticles are the same up to the opposite
signs: $\bar{\alpha}=-\alpha$.

Using the fact that the decay selects the spin projections parallel to
the decay analyzer $\hat{\bf n}$, one can obtain the double angular
distribution of the decay analyzers by inserting in Eq. (\ref{eq9})
the analyzing powers 
$\mbox{\boldmath $\zeta$}_1 = \alpha_1 {\bf n}^{(1)} $ and
$\mbox{\boldmath $\zeta$}_2 = \alpha_2 {\bf n}^{(2)} $
(see also \cite{l16}):
\begin{equation}
\label{eq17}
W(\Omega_1,\Omega_2) = \frac{1}{(4\pi)^2} [ 1 +
\alpha_1 {\bf P_1} {\bf n}^{(1)} + \alpha_2 {\bf P_2}{\bf n}^{(2)}+
\alpha_1\alpha_2\sum_{i=1}^3 \sum_{k=1}^3 T_{ik} n_i^{(1)} n_k^{(2)}]. 
\end{equation} 

Integrating Eq. (\ref{eq17}) over all angles except the 
angle $\theta_{12}$ between the decay analyzers ${\bf n}^{(1)}$ and 
${\bf n}^{(2)}$, one gets (see also \cite{l15,l16}):
\begin{equation}
\label{eq19}
W(x) = \frac{1}{2} [ 1 + \frac13 {\rm tr}T\alpha_1\alpha_2 x],
\end{equation}
where $x\equiv \cos\theta_{12}= {\bf n}^{(1)}{\bf n}^{(2)}$ and the
trace of the correlation tensor ${\rm tr}T$ can be expressed through
the rotation invariant combination of the two--particle 
density matrix elements, such as the singlet or triplet fractions
$\rho_s\equiv \rho_0$ or $\rho_t\equiv \rho_1$,
$\rho_s+\rho_t=1$.
Using the fact that the eigen values of the operator
$\hat{\mbox{\boldmath $\sigma$}}^{(1)}\otimes 
\hat{\mbox{\boldmath $\sigma$}}^{(2)}$
are equal to -3 and 1 for the singlet and triplet states respectively,
one has
\begin{equation}
\label{eq20}
{\rm tr}T = \rho_t - 3\rho_s \equiv 4\rho_t -3.
\end{equation}
Particularly, in agreement with Eqs. (\ref{eq23}) and (\ref{eq29a}),
the correlations between the decay analyzers in the 
pure singlet and triplet states are of opposite signs and differ
in magnitude by a factor of three:
${\rm tr}T^s= -3{\rm tr}T^t= -3$.

The polarization vectors and the 
correlation tensor can be
easily determined experimentally by the method of moments.
For example, using Eq. (\ref{eq17}), one can calculate the
diagonal components of the correlation tensor 
according to the expressions:
\begin{equation}
\label{eq50a}
\begin{array}{c}
T_{11} = \frac{9}{\alpha_1\alpha_2}\left\langle \sin\theta_1
\cos\phi_1 \sin\theta_2 \cos\phi_2 \right \rangle;\quad
T_{22} = \frac{9}{\alpha_1\alpha_2}\left\langle \sin\theta_1 \sin
\phi_1 \sin\theta_2 \sin \phi_2 \right\rangle;
\\ \\
 T_{33} = \frac{9}{\alpha_1\alpha_2}\left\langle \cos\theta_1
\cos\theta_2\right \rangle,
\end{array}
\end{equation}
where $\theta_l$ and $\phi_l$ denote respectively
the polar and azimuthal angles of the decay analyzers
${\bf n}^{(l)}$, $l=1,2$.
One can directly calculate the sums (see also Eq. (\ref{eq19})):
\begin{equation}
\label{eq50b}
T_{11} +
T_{22} = \frac{9}{\alpha_1\alpha_2}\left\langle \sin\theta_1 \sin\theta_2 
\cos \phi_{12} \right\rangle;~~
{\rm tr} T = \frac{9}{\alpha_1\alpha_2}\left\langle \cos\theta_{12}
\right \rangle,
\end{equation}
where $\phi_{12}=\phi_{1}-\phi_{2}$ and $\cos\theta_{12}={\bf n}^{(1)}
{\bf n}^{(2)}$.

\section{Incoherent properties of the correlation tensor}
Let us consider the incoherent mixture of
factorizable two-particle spin states. In this case the
two-particle density matrix is a sum of the direct
products of one-particle density matrices only, with nonnegative
coefficients:
\begin{equation}
\label{eq39}
\hat{\rho}^{(1,2)} = \sum_{\{s\}} \sum_{\{t\}} b_{\{s,t\}}
\hat{\rho}_{\{s\}}^{(1)} \otimes \hat{\rho}_{\{t\}}^{(2)},
 \quad
 b_{\{s,t\}} \ge 0,\quad  \sum_{\{s\}} \sum_{\{t\}}b_{\{s,t\}} = 1.
\end{equation}
Inserting the one-particle density matrices for spin--1/2 particles 
\begin{equation}
\label{eq40}
\hat{\rho}_{\{s\}}^{(1)} = \frac{1}{2} (\hat{I}^{(1)} +
{\bf P}_{\{s\}}^{(1)}\hat{\mbox{\boldmath $\sigma$}}^{(1)}),\quad
\hat{\rho}_{\{t\}}^{(2)} = \frac{1}{2} (\hat{I}^{(2)} +
{\bf P}_{\{t\}}^{(2)}\hat{\mbox{\boldmath $\sigma$}}^{(2)})
\end{equation}
into Eq. (\ref{eq39}) and comparing with the general Eq. (\ref{eq6}), 
one obtains the following expressions for the
polarization vectors and the correlation tensor:
\begin{equation}
\begin{array}{c}
\label{eq42}
{\bf P}^{(1)} =\sum_{\{s\}}\sum_{\{t\}} b_{\{s,t\}}{\bf P}_{\{s\}}^{(1)},
\quad
{\bf P}^{(2)} =\sum_{\{s\}}\sum_{\{t\}} b_{\{s,t\}}{\bf P}_{\{t\}}^{(2)},
\\ \\
T_{ik}= \sum_{\{s\}}\sum_{\{t\}} b_{\{s,t\}} P_{\{s\}i}^{(1)}
 P_{\{t\}k}^{(2)}.
\end{array}
\end{equation}
Since the magnitudes of the polarization vectors do not exceed unity,
$\mid {\bf P}_{\{s\}}^{(1)}\mid \le 1$,
$\mid {\bf P}_{\{t\}}^{(2)}\mid \le 1$,
it follows from Eq. (\ref{eq42}) that, in the case
of the incoherent mixture of the factorizable two-particle
states, the diagonal components of the tensor $T_{ik}$  
satisfy the inequalities:
\begin{equation}
\begin{array}{c}
\label{eq44}
 \mid T_{11} + T_{22} \mid  \le 1, \quad \mid T_{22} + T_{33} \mid \le 1,
 \quad \mid T_{33}+ T_{11} \mid \le 1 ,
\\ \\
\mid {\rm tr}T \mid = \mid T_{11} + T_{22} + T_{33} \mid \le 1.
\end{array}
\end{equation}

It should be emphasized that the derived inequalities are
related to the classical correlations at the probability
(not amplitude) level.
They simply reflect the fact that a weighted mean of the scalar
products of some vectors cannot exceed the same mean of the
corresponding products of vector modulae.
In quantum mechanics, when we deal with the nonfactorizable coherent
superpositions of two-particle states, the inequalities (\ref{eq44}) 
may be substantially violated.
Particularly, for a two--particle singlet state
(see Eqs. (\ref{eq23})):  
 $
  T_{11} + T_{22} = T_{11} + T_{33} = T_{22} + T_{33} = -2,~
{\rm tr}T = -3.
$
For a two--particle triplet state (see
Eqs. (\ref{eq29a}) and (\ref{eq29b})) 
the last of the inequalities (\ref{eq44}) is always satisfied:
${\rm tr}T = 1$, while one of 
the other inequalities 
may be violated. Thus, $T_{11} + T_{22} \equiv 2w_0 > 1$ provided that
$w_0>1/2$;\footnote{
Note that in case of a pure triplet state (a state described by the wave 
function), one can always achieve this condition (except for a completely 
polarized state $P=1$) by a rotation which maximizes the
probability of zero spin projection to a value
$w_0^{\max}=(1+\sqrt{1-P^2})/2$.
}
since then $2|\Re \rho_{+-}|<1/2$, the remaining two
inequalities are satisfied. On the other hand, for $w_0<1/2$,
the violation of one of the 
inequalities $T_{11} + T_{33}= 1-w_0+2\Re \rho_{+-}\le 1$ or
$T_{22} + T_{33}= 1-w_0-2\Re \rho_{+-}\le 1$  may happen; the
required condition is $2|\Re \rho_{+-}|>w_{0}$.

\section{Bell inequalities}
The Bell inequalities \cite{l10} were
obtained at the probability level in the framework of the concept of
hidden parameters related to the common past of particles separated
from each other in space during the detection;  thus, the
coherent properties of the quantum-mechanical superpositions of
two-particle states were not taken into consideration. One of these
inequalities, as applied to particles with spin 1/2, has the form
\cite{l12}
\begin{equation}
\label{eq47}
\begin{array}{c}
Q= \mid\langle (\hat{\mbox{\boldmath $\sigma$}}^{(1)} {\bf n}) \otimes
(\hat{\mbox{\boldmath $\sigma$}}^{(2)} {\bf m}) \rangle +\langle
(\hat{\mbox{\boldmath $\sigma$}}^{(1)} {\bf n}) 
\otimes (\hat{\mbox{\boldmath $\sigma$}}^{(2)}
{\bf m}') \rangle +
\\ \\
+  \langle (\hat{\mbox{\boldmath $\sigma$}}^{(1)} {\bf n}'
\otimes (\hat{\mbox{\boldmath $\sigma$}}^{(2)} {\bf m}) \rangle - \langle
(\hat{\mbox{\boldmath $\sigma$}}^{(1)} {\bf n}') 
\otimes (\hat{\mbox{\boldmath $\sigma$}}^{(2)}
 {\bf m}') \rangle \mid\le 2,
\end{array}
\end{equation}
where ${\bf n}$, ${\bf m}$, ${\bf n}'$ and ${\bf m}'$ are arbitrary
unit vectors and
\begin{equation}
\label{eq46}
\langle (\hat{\mbox{\boldmath $\sigma$}}^{(1)}
{\bf n}) \otimes (\hat{\mbox{\boldmath $\sigma$}}^{(2)}{\bf m}) \rangle 
= \sum_i \sum_k T_{ik} n_i m_k
\end{equation}
is the average product of the double spin projections of the
first and second particle onto different axes ${\bf n}$ and
${\bf m}$.
The latter can be determined experimentally from the
double distribution of the directions of the decay analyzers
${\bf n}={\bf n}^{(1)}$ and ${\bf m}={\bf n}^{(2)}$
using Eq.(\ref{eq17}).
Selecting the number of
pairs of the decay analyzers 
$\Delta N({\bf n},{\bf m})$
in sufficiently narrow
intervals of the solid angles 
$\Delta\Omega_{{\bf n}}$ and $\Delta \Omega_{{\bf m}}$,
and denoting 
$
W_{\Delta}({\bf n}, {\bf m}) =
\Delta N({\bf n},{\bf m})/[N\Delta
\Omega_{{\bf n}}\Delta\Omega_{{\bf m}}],
$
where $N$ is the total number of pairs,
one can calculate
\begin{equation}
\label{eq50}
\langle (\mbox{\boldmath $\sigma$}^{(1)}{\bf n}) \otimes
(\mbox{\boldmath $\sigma$}^{(2)}{\bf m}) \rangle
= \frac{1}{\alpha_1\alpha_2}\left\{8\pi^2[W_{\Delta}({\bf n},
{\bf m}) + W_{\Delta}(-{\bf n}, - {\bf m})] - 1\right\}.
\end{equation}

It can be shown that the inequality (\ref{eq47}) holds
for the incoherent mixture of
factorizable two-particle states with the density matrix
defined according to Eqs. (\ref{eq39}). Indeed, taking into account
Eqs. (\ref{eq42}) for the corresponding correlation tensor 
and the relations
$
{\bf m} + {\bf m}' = 2 {\bf l} \cos (\beta/2),~
{\bf m} - {\bf m}' = 2 {\bf l}\,' \sin (\beta/2),
$
where ${\bf l}$ and ${\bf l}\,'$ are the mutually perpendicular unit
vectors, $\beta$ is the angle between the vectors ${\bf m}$ and
${\bf m}'$, the quantity $Q$ in Eq. (\ref{eq47}) can be rewritten as
\begin{equation}
\begin{array}{c}
\label{eq47b}
Q = \mid \sum_{\{s\}} \sum_{\{t\}} b_{\{s,t\}}
[({\bf P}_{\{s\}}^{(1)} {\bf n}) ({\bf P}_{\{t\}}^{(2)}{\bf l})
\,2 \cos(\frac{\beta}{2}) +
({\bf P}_{\{t\}}^{(1)} {\bf n}\,') ({\bf P}_{\{t\}}^{(2)} {\bf l}\,')
\,2 \sin (\frac{\beta}{2})]\mid =
\\ \\
  = \mid \sum_{\{s\}} \sum_{\{t\}} b_{\{s,t\}} P_{\{t\}}^{(2) (\parallel)}
[({\bf P}_{\{s\}}^{(1)} {\bf n})\, 2\cos(\frac{\beta}{2}) \cos\alpha_{\{t\}} +
({\bf P}_{\{s\}}^{(1)} {\bf n}\,')\, 2 \sin(\frac{\beta}{2})
\sin\alpha_{\{t\}}] \mid.
\end{array}
\end{equation}
Here $P_{\{t\}}^{(2) (\parallel)} $ is the projection of the vector
${\bf P}_{\{t\}}^{(2)}$ onto the plane (${\bf l},{\bf l}\,'$),
$\alpha_{\{t\}}$ is the angle between this projection and the vector
${\bf l}$. 
Since the magnitudes of the polarization vectors ${\bf P}_{\{s\}}^{(1)}$
and ${\bf P}_{\{t\}}^{(2)}$, as well as the magnitudes of their projections,
cannot exceed unity, one finally proves the inequality (\ref{eq47}):
\begin{equation}
\label{eq47c}
Q \le \sum_{\{s\}}\sum_{\{t\}} 2 b_{\{s,t\}} 
\max\mid \cos(\alpha_{\{t\}} \mp
\frac{\beta}{2}) \mid \le 2 \sum_{\{s\}}\sum_{\{t}\} b_{\{s,t\}}= 2.
\end{equation}

Similar to the inequalities (\ref{eq44}),
the Bell inequality (\ref{eq47}) can be violated for coherent superpositions
of two--particle states.
In particular, for  the singlet state, in accordance with Eqs.
(\ref{eq23}) and (\ref{eq46}) one has
$
 \langle  (\hat{\mbox{\boldmath $\sigma$}}^{(1)}{\bf n}) \otimes
(\hat{\mbox{\boldmath $\sigma$}}^{(2)}{\bf m}) \rangle = -
{\bf n}{\bf m},
$
so that the quantity
\begin{equation}
\label{eq48}
 Q = \mid -{\bf n}{\bf m} - {\bf n}{\bf m}' -{\bf n}'{\bf m} +
{\bf n}'{\bf m}'\mid.
\end{equation}
As a result, the maximal possible violation
of the Bell inequality: 
$
Q_{max} = 2\sqrt 2  > 2 
$
corresponds to the situation when the unit
vectors are selected in the same plane and satisfy the conditions
$
{\bf n} \perp {\bf n}',~{\bf m} \perp {\bf m}',~
{\bf n}{\bf m} = {\bf n}{\bf m}' = {\bf n}'{\bf m} =
\cos(\pi/4) = 1/\sqrt 2,~
 {\bf n}'{\bf m}' = \cos(3\pi/4) = - 1/\sqrt 2.
$

The analogous violations of the Bell inequality take place
also for the triplet state with the zero projection onto
any axis $z$, provided the vectors ${\bf n}$, ${\bf m}$,
${\bf n}'$ and ${\bf m}'$ are selected in the plane
$(x,y)$. For such vectors (see Eq. (\ref{eq29}) with $w_0=1$)
$
  \langle (\mbox{\boldmath $\sigma$}^{(1)} {\bf n}) \otimes
(\mbox{\boldmath $\sigma$}^{(2)} {\bf m}) \rangle = + {\bf n}{\bf m} ,
$
so that the quantity $Q$ coincides with the singlet expression
in Eq. (\ref{eq48}). In the general aligned 
triplet state (a state with the diagonal density matrix),
the double spin correlation
$\langle (\mbox{\boldmath $\sigma$}^{(1)} {\bf n}) \otimes
(\mbox{\boldmath $\sigma$}^{(2)} {\bf m}) \rangle $
in the plane $(x,y)$ and the corresponding quantity $Q$ 
scale with the probability $w_0$ of the zero spin
projection onto the $z$--axis.
The Bell inequality (\ref{eq47}) can then be violated for $w_0>1/\sqrt{2}$
only. Recall that the inequality $|T_{11}+T_{22}|\le 1$
appears to be more stringent, being always violated for $w_0>1/2$.

It should be stressed that there exists the difference of
principle between the singlet state in quantum mechanics
and the incoherent mixture of the products of two one-particle
states with opposite projections onto the isotropically distributed axes.
In the latter case
$
T_{ik} = - \delta_{ik}/3,~
\langle (\hat{\mbox{\boldmath $\sigma$}}^{(1)}{\bf n}) \otimes
(\hat{\mbox{\boldmath $\sigma$}}^{(2)}{\bf m}) \rangle = -
{\bf n} {\bf m}/3,
$
so that the inequalities (\ref{eq44}), as well as 
the Bell inequality (\ref{eq47}),
are valid.

\section{Spin correlations} 

Consider, for example, the processes 
$e^-e^+, q \bar{q}\rightarrow \mu^-\mu^+$
well below the $Z^0$ threshold or the process
$q \bar{q}\rightarrow t \bar{t}$.
Due to parity conservation, the corresponding dominant tree diagrams
(the s-channel photon or gluon exchange) select 
the final particles in a triplet state, i.e. in a state with correlated
particle spins.\footnote{
The processes $q \bar{q}\rightarrow \mu^-\mu^+, t \bar{t}$
are relevant for the production of Drell-Yan dimuons and
top-quark pairs in hadronic collisions.
The contribution of the competing process 
$g g \rightarrow t \bar{t}$ is small at Tevatron, it will however 
dominate at LHC.
Note that the spin composition of the $t\bar{t}$-pair in this process
changes with the increasing energy from the singlet (due to the Landau-Yang
theorem forbidding the total two--gluon angular momentum $J=1$) \cite{har91}
to the triplet (due to the helicity conserving gluon coupling to the
relativistic quarks). 
} 
Directing the z-axis parallel to the production plane normal and the
y-axis - antiparallel to the muon (top-quark) c.m.s. velocity vector
$\mbox{\boldmath $\beta$}$,
the nonzero components of the correlation tensor in the tree
approximation become
(see, e.g., \cite{bra96}):
\begin{equation}
\label{eq*}
\begin{array}{c}
T_{11}=A^{-1}(2-\beta^2)\sin^2\theta,~~
T_{22}=A^{-1}(2\cos^2\theta+\beta^2\sin^2\theta),~~
T_{33}=-A^{-1}\beta^2\sin^2\theta,
\\ \\
T_{12}=T_{21}=-A^{-1}\gamma^{-1}\sin{2\theta},~~
A=2-\beta^2\sin^2\theta,
\end{array}
\end{equation}
where $\theta$, $\beta$ and $\gamma=(1-\beta^2)^{-1/2}$
are the c.m.s. muon (top-quark) production angle, velocity and
Lorentz factor.
Using the transformation properties of the correlation tensor under
the rotations of the coordinate system, 
one can easily prove that the choice of the quantization axis parallel
to the production plane normal maximizes the probability of zero spin
projection: $w_0^{\max}=(2-\beta^2\sin^2\theta)^{-1}$.
Since for a triplet state $T_{11}+T_{22}=2w_0$, the first of the
inequalities (\ref{eq44}) will be violated provided 
$\beta\sin\theta \ne 0$.

It should be noted that near threshold, the considered processes are
strongly influenced by the Coulomb or colour-Coulomb final state
interaction (FSI). Taking into account the point-like character of
these processes and neglecting the effect of 
finite lifetimes of produced particles
(see, however, \cite{fad88}) 
the FSI can be approximately taken into account by multiplying the
production amplitudes by the stationary solutions of the Coulomb 
scattering problem at zero separation:
$\psi_{{\bf k}^*}^{c(-)}(0)=e^{-i\delta_c}A_c^{1/2}$,
where ${\bf k}^*$ is the c.m.s. muon (top-quark) 3-momentum,
$\delta_c=\arg \Gamma(1+i\eta)$ is the Coulomb s-wave shift,
$A_c=2\pi\eta/[\exp(2\pi\eta)-1]$ is the Coulomb penetration factor,
$\eta=(ak^*)^{-1}$, $a=(-\alpha\mu)^{-1}$ is the Bohr radius
(taken negative in case of attraction),
$\mu$ is the pair reduced mass and $\alpha$ is here the fine structure
constant for $\mu^-\mu^+$-pair while
$\alpha=\frac43 \alpha_s$ ($-\frac16 \alpha_s$) for colour singlet (octet) 
$t \bar{t}$-pair \cite{har91}.
It is important that this interaction, being spin-independent,
has no influence on the spin structure of the amplitudes
of the considered processes.
Particularly, it leaves unchanged  
the components of the correlation tensor in Eqs. (\ref{eq*}).

For identical particles, independent of the production dynamics,
the effect of Bose or Fermi quantum statistics (QS)
leads to the spin correlation 
at small relative momenta
$Q=2k^*$ in the two--particle rest frame. This is obvious
due to the fact that the total spin $S$ of
two identical particles and the orbital angular momentum $L$ in their
c.m.s. satisfy the well-known equality 
\cite{l20}:
$
(-1)^{S + L} =1.
$
When the momentum difference $Q$ approaches zero, the states with nonzero
orbital angular momenta disappear, and only those with
$L=0$ and even total spin $S$ remain. As a result, 
at $Q \rightarrow 0$, two
identical spin--1/2 particles (e.g., two
protons or two $\Lambda$-particles) can be produced only in the
singlet state \cite{l6}. This conclusion is clearly model independent.
The corresponding width of the singlet enhancement or triplet
suppression is inversely related to the effective radius $r_0$
of the emission region \cite{l15,l16}.
The measurement of the singlet or triplet fractions $\rho_s$ or
$\rho_t$ thus yields similar though completely independent 
information on the space--time
separation of the produced particles as the standard correlation
femtoscopy technique. The latter exploits the fact that the
correlation function at small $Q$ is sensitive to particle separation
due to the effects of QS, Coulomb and strong FSI. 
Note that the
correlation function can be defined as a ratio $R(p_1,p_2)$ of the
two--particle production cross section to the reference one which
would be observed in the absence of the effects of QS and FSI.
The reference distribution is usually constructed by mixing the
particles from different events.

In the model of independent one--particle sources, the
effects of QS and FSI are taken into account merely multiplying
the initial amplitude for a total spin $S$ by 
a properly symmetrized stationary solution of the scattering problem
$\psi_{{\bf k}^{*}}^{S(-)}({\bf r}^{*})=
[\psi_{-{\bf k}^{*}}^{S(+)}({\bf r}^{*})]^*\equiv \psi^{S*}$, where
${\bf k}^{*}= {\bf p}^{*}_{1} = -{\bf p}^{*}_{2}$ 
and ${\bf r}^{*}= {\bf r}^{*}_{1} -{\bf r}^{*}_{2}$ 
(the minus sign of the vector ${\bf k}^{*}$ 
in $\psi_{-{\bf k}^{*}}^{S(+)}({\bf r}^{*})$ corresponds to the reverse 
in time direction of the emission process).
Particularly, one has \cite{l22,l23}:
$
R(p_{1},p_{2})= 
\sum_{S}\widetilde{\rho}_{S} 
\langle |\psi_{-{\bf k}^{*}}^{S(+)}({\bf r}^{*})|^{2} 
\rangle _{S} 
\equiv \sum_{S} R_S(p_{1},p_{2}) 
$
and
$
\rho_S= \widetilde{\rho}_S
\langle|\psi^S|^2\rangle/R\equiv R_S/R.
$ 
The averaging is done over the emission points 
of the two particles in a state with total spin $S$ 
populated in the absence of QS and FSI with the probability 
$\widetilde{\rho}_{S}$, 
$\sum_{S}\widetilde{\rho}_{S} = 1$.
For spin--1/2 particles initially emitted independently
with the polarizations $\widetilde{\bf P}_1$ and 
$\widetilde{\bf P}_2$: 
$ 
\widetilde{\rho}_s=(1-\widetilde{\bf P}_1\cdot\widetilde{\bf P}_2)/4,~
\widetilde{\rho}_t=(3+\widetilde{\bf P}_1\cdot\widetilde{\bf P}_2)/4. 
$ 
The expressions for the components of the correlation tensor and particle
polarizations are rather lengthy and can be found in ref. \cite{l16}.
Here we present only the simple result for the case of low--$Q$
pairs of identical spin--1/2 particles when one can put
$\widetilde{\bf P}_1\doteq \widetilde{\bf P}_2\doteq \widetilde{\bf P}$.
Then 
\begin{equation}  
{\bf P} =  
\widetilde{\bf P}\rho_t/\widetilde{\rho}_t,~~~
T_{ik} = 
\widetilde{P}_{i}\widetilde{P}_{k}\rho_t/\widetilde{\rho}_t 
+\delta_{ik}\left(\rho_t/\widetilde{\rho}_t-1\right). 
\label{PT3}  
\end{equation} 

It may be seen from Eqs. (\ref{PT3}) that, 
in the presence of QS and FSI, 
the spins of the 
initially unpolarized ($\widetilde{P}_{1i}=\widetilde{P}_{2i}=0$) 
and uncorrelated ($\widetilde{T}_{ik}=0$) 
particles remain unpolarized but not uncorrelated: 
${T}_{ik}= -\delta_{ik} 
\langle|\psi^s|^2-|\psi^t|^2\rangle/ 
\langle|\psi^s|^2+3|\psi^t|^2\rangle$. 
For $\psi^t=0$ ($\psi^s=0$) the latter tensor reduces to 
a pure singlet (triplet) one: $T_{ik}^s=-\delta_{ik}$ 
($T_{ik}^t=\frac13\delta_{ik}$). 
On the other hand, for initially polarized identical 
particles the polarization vectors vanish and 
${T}_{ik}\rightarrow T_{ik}^s$ 
at $Q\rightarrow 0$ due to forbidden triplet amplitude $\psi^t$ at 
$Q=0$.

It should be emphasized that the correlation of the polarizations
of two spin--1/2 particles, conditioned by their identity, is
maximal for $ Q \rightarrow 0 $ ($ {\rm tr}T \rightarrow -3$ independently
of the value of the initial polarization 
$\widetilde{\bf P}$ - the singlet state) . 
At sufficiently large $Q$, the model of
independent one-particle sources yields 
$T_{ik}= \widetilde{P}_i\widetilde{P}_k$ so that the
spin correlations vanish for unpolarized particles  
($ {\rm tr}T \rightarrow 0$).

Note that recent measurements of 
$\Lambda \Lambda$ correlations in multihadronic $Z^0$ decays
at LEP point to rather small effective radius of the
$\Lambda$ emission region: 
$r_0 \sim 0.1-0.2$ fm \cite{l24}. One can therefore expect that 
at $Q < 1/r_0 \sim 1-2$ GeV/c the system of two $\Lambda$-hyperons 
is created mainly in the singlet state.\footnote{
A large statistics of $\Lambda \Lambda$ pairs is also available
in heavy ion collisions at SPS and soon will be accumulated at RHIC.
However, due to much larger effective radius $r_0$ of
several fm, the singlet state dominates here only in quite narrow
region of $Q < 0.1$ GeV/c.
}   
For such systems, the inequalities
(\ref{eq44}) for the sums of the components of the correlation tensor
and the Bell inequality (\ref{eq47}) could then be violated.
Indeed, the expected suppression of the triplet fraction at small Q
was observed in several LEP experiments \cite{l24,l24a}.
Particularly, the ALEPH result:
$\rho_t=0.36\pm 0.30\pm 0.08$ at $Q= 0-1.5$ GeV/c
indicates the violation of the inequality
$|{\rm tr}T|\equiv |4\rho_t-3| \le 1$.

\section{Conclusions}
We have performed the theoretical
analysis of spin correlations in the system of two
spin--1/2 particles using their asymmetric (parity violating) decays.
It is shown that the spin correlation tensor can be determined from
the angular correlations of the decay analyzers, 
particularly, for two $\Lambda$--particles both decaying into the channel
$\Lambda \rightarrow p\pi^-$, - from the correlations
of the directions of the decay protons
in the respective $\Lambda$--rest frames.

We have derived the inequalities, including those of Bell, 
for linear combinations of the
components of the spin correlation tensor valid in the case of incoherent
mixture of two-particle factorizable spin states.
The violation of these inequalities is connected with the general 
quantum-mechanical effect (first considered by Einstein, Podolsky and Rosen)
and can serve as a crucial test of the basic principles of quantum mechanics.

We have considered some examples of the processes allowing one to
verify the consequences of the quantum--mechanical coherence
with the help of the two--particle spin correlations 
measured in asymmetric particle decays, 
the coherence arising either due to the production dynamics 
(dominant triplet states in the processes 
$e^-e^+, q \bar{q}\rightarrow \mu^-\mu^+, t \bar{t}$)
or due to the effect of quantum
statistics at small relative momenta
(dominant singlet $\Lambda\Lambda$-state). 
 
\section*{Acknowledgments} 
This work was supported by GA Czech Republic, 
Grant No. 202/01/0779.

\end{document}